\documentclass[prd,aps,showpacs,epsf,floats,onecolumn]{revtex4}%
\usepackage{amssymb}
\usepackage{amsfonts}
\usepackage{amsmath}
\usepackage{graphicx}%
\providecommand{\U}[1]{\protect\rule{.1in}{.1in}}

\begin{document}
\title{\textbf{An information geometric perspective on the complexity of macroscopic
predictions arising from incomplete information}}
\author{\textbf{Sean Alan Ali}$^{1}$, \textbf{Carlo Cafaro}$^{2}$, \textbf{Steven
Gassner}$^{3}$, \textbf{Adom Giffin}$^{4}$}
\affiliation{$^{1}$Albany College of Pharmacy and Health Sciences, 12208 Albany, New York, USA}
\affiliation{$^{2,3}$SUNY Polytechnic Institute, 12203 Albany, New York, USA}
\affiliation{$^{4}$Clarkson University, 13699 Potsdam, New York, USA}

\begin{abstract}
Motivated by the presence of deep connections among dynamical equations,
experimental data, physical systems, and statistical modeling, we report on a
series of findings uncovered by the Authors and collaborators during the last
decade within the framework of the so-called Information Geometric Approach to
Chaos (IGAC). The IGAC is a theoretical modeling scheme that combines methods
of information geometry with inductive inference techniques to furnish
probabilistic descriptions of complex systems in presence of limited
information. In addition to relying on curvature and Jacobi field
computations, a suitable indicator of complexity within the IGAC framework is
given by the so-called Information Geometric Entropy (IGE). The IGE\ is an
information geometric measure of complexity of geodesic paths on curved
statistical manifolds underlying the entropic dynamics of systems specified in
terms of probability distributions. In this manuscript, we discuss several
illustrative examples wherein our modeling scheme is employed to infer
macroscopic predictions when only partial knowledge of the microscopic nature
of a given system is available. Finally, we include comments on the strengths
and weaknesses of the current version of our proposed theoretical scheme in
our concluding remarks.

\end{abstract}

\pacs{Chaos (05.45.-a), Complexity (89.70.Eg), Entropy (89.70.Cf), Inference Methods
(02.50.Tt), Information Theory (89.70.+c), Probability Theory (02.50.Cw),
Riemannian Geometry (02.40.Ky).}
\maketitle

\pagebreak

\section{Introductory background}

Characterizing and to some degree understanding the emergence and evolutionary
development of biological systems represent one of the most compelling
motivations to investigate the highly elusive concept of complexity
\cite{landauer88,bennett89,gellmann95,feldman98,adami02}.

Entropic inference methods \cite{caticha12} have recently\textbf{ }been
combined with information geometric techniques \cite{amari,amari2,amari3} in
an effort to uncover quantitative indicators of complexity suitable for
application to statistical models used to render probabilistic descriptions of
systems about which only limited information is known. Within this approach,
the notion of complexity associated with statistical models can be regarded as
a measure of the difficulty of inferring macroscopic predictions due to the
inherent lack of complete knowledge about the microscopic degrees of freedom
of the system being investigated.

This line of research was initially referred to as the \emph{Information
Geometric Approach to Chaos }(IGAC) \cite{cafarophd}. A schematic outline of
the IGAC\ theoretical framework is presented as follows: once the microscopic
degrees of freedom of a complex system of arbitrary nature are identified and
its relevant information constraints are chosen, entropic methods are used to
establish an initial, static statistical model of the complex system. The
statistical model describing the complex system is defined by means of
probability distributions parametrized in terms of statistical macrovariables
which, in turn, depend upon the specific functional form of the information
constraints assumed to be relevant for the implementation of statistical
inferences. The next step in the program concerns the evolution of the complex
system. In particular, assuming the complex system evolves, the evolution of
the associated statistical model from its initial to final configurations is
determined by means of the so-called Entropic Dynamics (ED, \cite{catichaED}).
Entropic Dynamics constitute a variant of information-constrained dynamics
that is constructed on statistical manifolds whose elements correspond to
probability distributions. Furthermore, these distributions are in one-to-one
relation with a suitable set of statistical macrovariables that define a
parameter space which serves to provide a convenient parametrization of points
on the original statistical manifold. The ED framework prescribes the
evolution of probability distributions by means of an entropic inference
principle: in particular,\textbf{ }starting from the initial configuration,
the motion toward the final configuration occurs via the maximization of the
logarithmic relative entropy (Maximum relative Entropy method- MrE method,
\cite{caticha12}) between any two consecutive intermediate configurations. At
this juncture, it is worth noting that ED\ provides only the \emph{expected},
but not the \emph{actual}, trajectories of the system. Inferences within ED
rely on the nature of the chosen information constraints that are utilized at
the level of\textbf{ }the MrE algorithm. The validation of modeling schemes of
this type can only be verified \emph{a posteriori}. If discrepancies occur
between experimental observations and the inferred predictions, a new set of
information constraints must be selected \cite{jaynes85,dewar09,giffin16}.
This is an extremely important feature of the MrE algorithm and was recently
reconsidered in Ref. \cite{cafaropre16}. The evolution of probability
distributions is specified in terms of the results of the maximization
procedure described above, namely a geodesic evolution of the statistical
macrovariables \cite{caticha12}. A measure of distance between two different
probability distributions is quantified by the Fisher-Rao information metric
\cite{amari}. This distance can be interpreted as the degree of
distinguishability between two distributions. After having determined the
information metric, one can readily apply the standard methods of Riemannian
differential geometry to investigate the geometric structure of the
statistical manifold underlying the entropic motion which characterizes the
evolution of the probability distributions. Generally speaking, conventional
Riemannian geometric quantities such as Christoffel connection coefficients of
the second kind, Ricci tensor, Riemannian curvature tensor, sectional
curvatures, scalar curvature, Weyl anisotropy tensor, Killing fields, and
Jacobi fields can be computed in the standard manner \cite{thorne73}. More
specifically, the chaoticity (i.e. temporal complexity) of such statistical
models can be investigated by means of suitably chosen measures, such as: the
signs of the scalar and sectional curvatures of the statistical manifold, the
asymptotic temporal behavior of Jacobi fields, the existence of Killing
vectors, and the existence of a non-vanishing Weyl anisotropy tensor. In
addition to these measures, complexity within the IGAC\ approach can also be
quantified\textbf{ }by means of the so-called information geometric entropy
(IGE), originally presented in \cite{cafarophd}.

For a review of the MrE inference algorithm, we refer to Ref.
\cite{caticha12,adom06,adom07,adomphd}. Furthermore, for a detailed
presentation on the ED approach used in this manuscript, we refer to Ref.
\cite{catichaED}. Finally, for the sake of brevity, we have omitted
mathematical details in our main presentation. For the sake of
self-consistency however, we have added useful mathematical details on the
notions of curvature, information geometric entropy, and Jacobi fields in
Appendices A, B, and C, respectively.

\section{Illustrative examples}

In this section, we outline all available applications concerning the
characterization of the complexity of geodesic paths on curved statistical
manifolds within the IGAC framework. For a conventional approach to the
Riemannian geometrization of classical Newtonian dynamics, we refer to Refs.
\cite{casetti96,dibari98,casetti00,pettini07}. We remark that in what follows
we only report on the asymptotic behavior of the chosen complexity indicators.
These complexity indicators are parameterized with respect to the statistical
affine parameter used to quantify temporal change in the geodesic analysis on
the underlying statistical manifold. Furthermore, the order in which these
illustrative examples are presented in this manuscript is chronological.

Very preliminary concepts and applications of the IGAC appeared originally in
Refs. \cite{cafaroaip06, cafaroaip07, cafaroaip08}. For a more recent review
on the IGAC and the IGE with more detailed\textbf{ }computations, we refer to
Refs. \cite{alips12, cafarobrescia13} and Ref. \cite{cafaroamc10}, respectively.

\subsection{Example 1: Uncorrelated Gaussian statistical models}

Previously, the IGAC modeling framework was used to investigate the
information geometric properties of a system of arbitrary nature containing
$l$ degrees of freedom, each one characterized by two pieces of relevant
information, namely its mean and its variance \cite{cafaropd07, cafaroijtp08}.
From an information geometric perspective, it was determined that the family
of statistical models corresponding to such a system is Gaussian in form. This
set of Gaussian distributions generate a non-maximally symmetric
$2l$-dimensional statistical manifold $\mathcal{M}_{s}$ exhibiting a constant
negative scalar curvature $\mathcal{R}_{\mathcal{M}_{s}}$ proportional to the
number of degrees of freedom of the system (Appendix A),%
\begin{equation}
\mathcal{R}_{\mathcal{M}_{s}}=-l\text{.} \label{curvature18}%
\end{equation}
From a dynamical standpoint, it was found that the system explores statistical
volume elements on $\mathcal{M}_{s}$ at an exponential rate. Specifically, in
the asymptotic limit, the IGE $\mathcal{S}_{\mathcal{M}_{s}}$\ increases
linearly with respect to the statistical affine parameter $\tau$ and is
proportional to the number of degrees of freedom $l$ according to (Appendix
B),%
\begin{equation}
\mathcal{S}_{\mathcal{M}_{s}}\left(  \tau\right)  \overset{\tau\rightarrow
\infty}{\sim}l\lambda\tau\text{,} \label{entropy18}%
\end{equation}
where $\lambda$ represents the maximal positive Lyapunov exponent
\cite{strogatz15} characterizing the statistical model. Furthermore, the
geodesic paths on $\mathcal{M}_{s}$ were shown to be hyperbolic trajectories.
Finally, by solving the geodesic deviation equations, it was determined that
in the asymptotic limit, the Jacobi vector field intensity $J_{\mathcal{M}%
_{s}}$ exhibits exponential divergence and is proportional to the number of
degrees of freedom $l$ (Appendix C),%
\begin{equation}
J_{\mathcal{M}_{s}}\left(  \tau\right)  \overset{\tau\rightarrow\infty}{\sim
}l\exp\left(  \lambda\tau\right)  \text{.} \label{jacobi18}%
\end{equation}
Since\textbf{ }the exponential divergence of the Jacobi vector field intensity
$J_{\mathcal{M}_{s}}$ is a known classical feature of chaos, from Eqs.
(\ref{curvature18}), (\ref{entropy18}), and (\ref{jacobi18}), the Authors
propose that $\mathcal{R}_{\mathcal{M}_{s}}$, $\mathcal{S}_{\mathcal{M}_{s}}$
and $J_{\mathcal{M}_{s}}$ each behave as proper indicators of chaoticity, with
each being proportional to the number of Gaussian-distributed microstates of
the system. Despite being verified in this special scenario, this
proportionality constitutes\textbf{ }the first known example appearing in the
literature of a possible substantive connection among information geometric
indicators of chaoticity deduced from the probabilistic descriptions of
dynamical systems.

\subsection{Example 2: Correlated Gaussian statistical models}

In \cite{aliphysica10}, the IGAC framework was applied to analyze the
information-constrained dynamics of a system with two correlated,
Gaussian-distributed microscopic degrees of freedom. As a working hypothesis,
the degrees of freedom were assumed to be characterized by the same variance.
The presence of microscopic correlations lead to the emergence of asymptotic
information geometric compression of the statistical macrostates explored by
the system at a faster rate than that observed in the absence of microscopic
correlations. In particular, it was found that in the asymptotic limit
(Appendix B)%
\begin{equation}
\left[  \exp(\mathcal{S}_{\mathcal{M}_{s}}\left(  \tau\right)  )\right]
_{\text{correlated}}\overset{\tau\rightarrow\infty}{\sim}\mathcal{F}\left(
\rho\right)  \cdot\left[  \exp(\mathcal{S}_{\mathcal{M}_{s}}\left(
\tau\right)  )\right]  _{\text{uncorrelated}}\text{,} \label{fuckyou}%
\end{equation}
where $\mathcal{F}\left(  \rho\right)  $ in Eq. (\ref{fuckyou}) with\textbf{
}$0\leq$ $\mathcal{F}\left(  \rho\right)  \leq1$ is defined as,%
\begin{equation}
\mathcal{F}\left(  \rho\right)  \overset{\text{def}}{=}\frac{1}{2^{\frac{5}%
{2}}}\left[  \sqrt{\frac{4\left(  4-\rho^{2}\right)  }{\left(  2-2\rho
^{2}\right)  ^{2}}}\left(  \frac{2+\rho}{4\left(  1-\rho^{2}\right)  }\right)
^{-\frac{3}{2}}\right]  \text{.}%
\end{equation}
The function $\mathcal{F}\left(  \rho\right)  $ is a monotonic decreasing
compression factor defined for any value of the correlation coefficient
$\rho\in\left(  0\text{, }1\right)  $. This result constitutes an explicit
connection between correlations at the \emph{microscopic} level and complexity
at the \emph{macroscopic} level \cite{lebowitz81,lebowitz93,lebowitz99}. This
result provides a concise and transparent description of the behavioral change
of the macroscopic complexity of a statistical model caused by the presence of
microscopic correlations.

\subsection{Example 3: Inverted harmonic oscillators}

In a broad sense, it is known that the primary issues addressed by the General
Theory of Relativity are twofold: first, one is interested in the manner that
space-time geometry evolves in response to mass-energy; second, one seeks to
understand how mass-energy configurations move in such a space-time\ geometry.
Within the IGAC approach, one focuses on how systems move in a given
statistical geometry, while the evolution of the statistical geometry itself
is neglected. The realization that there are two distinct and separate aspects
to this scenario served as a turning point in the development of the IGAC
framework that led to an intriguing result. The first formal result in this
research direction was presented in Ref. \cite{catichaaip07}\textbf{,} where
the possibility of exploiting well-established principles of inference to
derive Newtonian dynamics from relevant prior information codified in an
appropriate statistical manifold was explored. The key working hypothesis in
that derivation was the assumed existence of an irreducible uncertainty in the
location of particles which requires the state of a particle to be
characterized by a probability distribution. The corresponding configuration
space is therefore a curved statistical manifold whose Riemannian geometry is
defined by the Fisher-Rao information metric. The expected trajectory follows
from the MrE method viewed as a principle of inference. In this approach,
there is no need for additional physical postulates such as an action
principle or equation of motion, nor for the concept of mass, momentum, or of
phase space; not even the notion of external, absolute time. Newton's
mechanics for any number of particles interacting among themselves and with
external fields is completely reproduced by the resulting entropic dynamics.
Furthermore, both the interactions between particles and their masses are
explained as a consequence of the underlying statistical manifold.

From a more applied perspective, building upon the results found in
\cite{catichaaip07}, an information geometric analogue of the Zurek-Paz
quantum chaos criterion in the classical reversible limit
\cite{zurek93,zurek94,zurek95} was presented in
\cite{cafaroejtp08,cafarocsf09}. In these works, the IGAC\ framework was used
to investigate a set of\textbf{\ }$l$\textbf{ }three-dimensional, uncoupled,
anisotropic, inverted harmonic oscillators (IHO) characterized by an Ohmic
distributed frequency spectrum. Omitting technical details, in Refs.
\cite{cafaroejtp08,cafarocsf09} it was shown that the asymptotic temporal
behavior of the IGE for such a system is given by (Appendix B),%
\begin{equation}
\mathcal{S}_{\mathcal{M}_{\text{IHO}}^{\text{(}l\text{)}}}\left(  \tau\text{;
}\omega_{1}\text{,\ldots, }\omega_{l}\right)  \overset{\tau\rightarrow\infty
}{\sim}\Omega\tau\text{,} \label{Fin}%
\end{equation}
where,
\begin{equation}
\Omega\overset{\text{def}}{=}\overset{l}{\underset{i=1}{\sum}}\omega
_{i}\text{,}%
\end{equation}
and $\omega_{i}$ with $1\leq i\leq l$ denotes the frequency of the $i$-th
inverted harmonic oscillator. Equation (\ref{Fin}) displays an asymptotic,
linear IGE growth for the set of inverted harmonic oscillators and can be
viewed as an extension of the result of Zurek and Paz presented in Refs.
\cite{zurek94,zurek95} to an ensemble of anisotropic, inverted harmonic
oscillators within the IGAC framework. Specifically, in \cite{zurek94,zurek95}%
, Zurek and Paz investigated the effects of decoherence in quantum chaos by
considering a single unstable harmonic oscillator with frequency $\Omega$
characterized by a potential $V\left(  x\right)  $,%
\begin{equation}
V\left(  x\right)  \overset{\text{def}}{=}-\frac{\Omega^{2}x^{2}}{2}\text{,}
\label{vimo}%
\end{equation}
coupled to an external environment. They determined that in the reversible
classical limit, the von Neumann entropy of such a system increases linearly
at a rate determined by the Lyapunov exponent $\Omega$ according to,
\begin{equation}
\mathcal{S}_{\text{quantum}}^{\text{(chaotic)}}\left(  \tau\right)
\overset{\tau\rightarrow\infty}{\sim}\Omega\tau\text{.} \label{hud1}%
\end{equation}
In \cite{cafaroejtp08,cafarocsf09}, Eq. (\ref{Fin}) was essentially proposed
as the classical information geometric analog of Eq. (\ref{hud1}).

\subsection{Example 4: Quantum spin chains}

We recall that it is commonly conjectured that spectral correlations of
classically integrable systems are well described by Poisson statistics and
that quantum spectra of classically chaotic systems are universally correlated
according to Wigner-Dyson statistics. The former and the latter conjectures
are known as the BGS (Bohigas-Giannoni-Schmit, \cite{bohigas84} and BTG
(Berry-Tabor-Gutzwiller, \cite{gutzwiller90}) conjectures, respectively.

In \cite{cafaromplb08,cafarophysica08}, the IGAC\ was employed to analyze the
entropic dynamics on statistical manifolds induced by classical probability
distributions commonly used in the investigation of regular and chaotic
quantum energy level statistics. In particular, an information geometric
characterization of the chaotic (integrable) energy level statistics of a
quantum antiferromagnetic Ising spin chain immersed in a tilted (transverse)
external magnetic field was proposed. The IGAC of a Poisson distribution
coupled to an Exponential bath (representing a spin chain in a
\textit{transverse} magnetic field, corresponding to the integrable case) and
that of a Wigner-Dyson distribution coupled to a Gaussian bath (representing
a\textbf{ }spin chain in a \textit{tilted} magnetic field, corresponding to
the chaotic case) were studied. Remarkably, it was found that in the former
case, the IGE exhibits asymptotic logarithmic growth (Appendix B),%
\begin{equation}
\mathcal{S}_{\mathcal{M}_{s}}^{\text{(integrable)}}\left(  \tau\right)
\overset{\tau\rightarrow\infty}{\sim}c\log\left(  \tau\right)  +\tilde
{c}\text{,} \label{sintegrable}%
\end{equation}
while in the latter case, the IGE exhibits asymptotic linear growth (Appendix
B),%
\begin{equation}
\mathcal{S}_{\mathcal{M}_{s}}^{\text{(chaotic)}}\left(  \tau\right)
\overset{\tau\rightarrow\infty}{\sim}\mathcal{K}\tau\text{.} \label{schaos}%
\end{equation}
In this manuscript, $\log$ denotes the natural logarithmic function. The
quantities $c$ and $\tilde{c}$ in Eq. (\ref{sintegrable}) are two suitable
constants of integration that depend on the dimensionality of the underlying
statistical manifold and the boundary conditions on the statistical variables,
respectively. The quantity $\mathcal{K}$ in Eq. (\ref{schaos}) is a model
parameter that describes the asymptotic temporal rate of change of the IGE. In
view of these findings, it was conjectured that the IGAC framework may be of
some utility when analyzing\textbf{ }potential physical applications in the
field of quantum energy level statistics. In such cases, the IGE\ would serve
the role of the entanglement entropy defined in standard quantum information
theory \cite{prosenpre07,prosenpra07}.

\subsection{Example 5: Statistical embedding and complexity reduction}

Reducing the complexity of statistical models is a very active field of
research \cite{transtrum15,daniels15,daniels15b}. Building upon the
exploratory analysis presented in Ref. \cite{cafarops10}, the IGAC approach
was utilized\textbf{ }in Ref. \cite{cafaropd11} to investigate the
$2l$-dimensional Gaussian statistical model $\mathcal{M}_{s}$ that is induced
by an appropriate embedding within a larger $4l$-dimensional Gaussian manifold
endowed with a Fisher-Rao information metric $g_{\mu\nu}$ containing
non-trivial off diagonal terms. These terms arise due to the presence of the
correlational structure coefficients $\rho_{k}$ with $1\leq k\leq l$ that
characterize the embedding constraints among the statistical variables on the
larger manifold. Two main findings were obtained. First, a power law decay of
the IGE (Appendix B) at a rate determined by the coefficients $\rho_{k}$ was
observed%
\begin{equation}
\mathcal{S}_{\mathcal{M}_{s}}\left(  \tau;l,\lambda_{k},\rho_{k}\right)
\overset{\tau\rightarrow\infty}{\sim}\log\left[  \Lambda\left(  \rho
_{k}\right)  +\frac{\tilde{\Lambda}\left(  \rho_{k},\lambda_{k}\right)  }%
{\tau}\right]  ^{l}\text{,} \label{peppino}%
\end{equation}
where $\lambda_{k}$ are suitable constants of integration that specify the
problem being investigated, while the functional forms of the bounded
functions $\Lambda\left(  \rho_{k}\right)  $ and $\tilde{\Lambda}\left(
\rho_{k},\lambda_{k}\right)  $ can be found in Ref. \cite{cafaropd11}. Second,
in addition to Eq. (\ref{peppino}), it was found that the asymptotic
exponential divergence of the Jacobi vector field intensity was
attenuated.\textbf{ }Due to this observed attenuation, the Authors concluded
that the presence of such embedding constraints lead to the emergence of an
asymptotic information geometric compression of the explored macrostates on
the curved statistical manifold $\mathcal{M}_{s}$.\textbf{ }These results
serve as a further non-trivial step toward the characterization of the
complexity of microscopically correlated multidimensional Gaussian statistical
models of relevance in the mathematical modelling of realistic physical systems.

\subsection{Example 6: Entanglement induced via scattering}

Inspired by the preliminary analysis presented in \cite{kim11}, the IGAC was
used to provide an information geometric perspective of the quantum
entanglement generated by\ $s$-wave scattering \cite{law04} between two
Gaussian wave packets in Refs. \cite{kimpla11,kim12}. Within the IGAC
framework, the pre- and post-collisional quantum dynamical scenarios related
to an elastic, head-on collision were conjectured to be macroscopic
manifestations emerging from microscopic statistical structures. Exploiting
such a working hypothesis, the pre- and post-collisional scenarios were
described by uncorrelated and correlated Gaussian statistical models,
respectively. As a consequence, the Authors were capable of expressing the
entanglement strength in terms of scattering potential and incident particle
energies. Furthermore, the manner in which the entanglement duration is
related\textbf{\ }to the scattering potential and incident particle energies
was explained. Additionally, the connection between entanglement and
complexity of motion was discussed. In particular, it was shown that in the
asymptotic limit (Appendix B),%
\begin{equation}
\left[  \exp(\mathcal{S}_{\mathcal{M}_{s}}\left(  \tau\right)  )\right]
_{\text{correlated}}\overset{\tau\rightarrow\infty}{\sim}\mathcal{F}\left(
\rho\right)  \cdot\left[  \exp(\mathcal{S}_{\mathcal{M}_{s}}\left(
\tau\right)  )\right]  _{\text{uncorrelated}}\text{,} \label{family}%
\end{equation}
where $\mathcal{F}\left(  \rho\right)  $ in Eq. (\ref{family}) with
$0\leq\mathcal{F}\left(  \rho\right)  \leq1$ is defined as,%
\begin{equation}
\mathcal{F}\left(  \rho\right)  \overset{\text{def}}{=}\sqrt{\frac{1-\rho
}{1+\rho}}\text{.}%
\end{equation}
The function $\mathcal{F}\left(  \rho\right)  $ is a monotonic decreasing
compression factor for any value of the correlation coefficient $\rho
\in\left(  0\text{, }1\right)  $. The work presented in Refs.
\cite{kimpla11,kim12} represent significant progress toward the goal of
understanding the relationship between entanglement and statistical
microcorrelations on the one hand and the effect of microcorrelations on the
complexity of informational geodesic paths on the other. Finally, due to the
consequences arising from Eq. (\ref{family}), the IGAC framework was proposed
as a potentially suitable platform to establish a sound information geometric
interpretation of quantum entanglement, including its connection to complexity
of motion in general physical scenarios.

\subsection{Example 7: Softening of classical chaos by quantization}

Comparing classical and quantum chaoticity (i.e. temporal complexity) and
explaining the reason why the former is stronger than the latter is of great
theoretical interest \cite{caron1, caron2, kroger}. It is usually conjectured
that the weakness of quantum chaos may be due to the Heisenberg uncertainty
relation. Following the preliminary investigations undertaken in Refs.
\cite{cafaroaip12,aliaip12,giffinaip13}, the IGAC was used to study both the
information geometry and the entropic dynamics of a three-dimensional Gaussian
statistical model and a two-dimensional Gaussian statistical model in which
the latter is obtained from the former via introduction of a suitable
macroscopic information constraint,%
\begin{equation}
\sigma_{x}\sigma_{y}=\Sigma^{2}\text{,} \label{macro}%
\end{equation}
where $\Sigma^{2}\in%
\mathbb{R}
_{0}^{+}$. The quantities $x$ and $y$ denote the microscopic degrees of
freedom of the system whose probabilistic description is being investigated.
The relation (\ref{macro}) resembles the canonical quantum mechanical minimum
uncertainty relation \cite{peres95}. It was found that the complexity of the
two-dimensional Gaussian statistical model, quantified in terms of the IGE, is
softened with respect to the complexity of the three-dimensional Gaussian
statistical model (Appendix B),%
\begin{equation}
\mathcal{S}_{\mathcal{M}_{s}}^{\text{(}2D\text{)}}\left(  \tau\right)
\overset{\tau\rightarrow\infty}{\sim}\left(  \frac{\lambda_{2D}}{\lambda_{3D}%
}\right)  \cdot\mathcal{S}_{\mathcal{M}_{s}}^{\text{(}3D\text{)}}\left(
\tau\right)  \text{,} \label{soft}%
\end{equation}
where $\lambda_{2D}$ and $\lambda_{3D}$ are two positive model parameters
(satisfying $\lambda_{2D}\leq\lambda_{3D}$) that specify the asymptotic
temporal rates of change of the IGE in the two-dimensional and
three-dimensional scenarios, respectively. In view of the similarity between
the selected macroscopic information constraint on the variances and the
phase-space coarse-graining imposed by the Heisenberg uncertainty relations,
the Authors argued that their work may provide a possible avenue to explain
the phenomenon of classical chaos suppression under the operation of
quantization within a novel information geometric perspective. We remark that
similar investigations were carried out in Ref. \cite{giffinentropy13} where
the analysis presented in Ref. \cite{cafaroosid12} was extended to the case in
which, in addition to the macroscopic constraint in Eq. (\ref{macro}), the
microscopic degrees of freedom $x$ and $y$ of the system are also correlated.

\subsection{Example 8:\ Topologically distinct correlational structures}

In Ref. \cite{felice14}, the asymptotic behavior of the IGE (Appendix B) for
bivariate and trivariate Gaussian statistical models in both the absence and
presence of microcorrelations was investigated. For correlated cases, various
correlational structures among the microscopic degrees of freedom of the
system were considered. It was determined that the complexity of macroscopic
inferences depends not only upon the amount of available microscopic
information, but also on the manner in which such microscopic information is
correlated. In particular, for a trivariate statistical model with two
correlations among the three degrees of freedom of the system (referred to as
the mildly connected case), it was found that in the asymptotic limit,
\begin{equation}
\left(  \exp\left[  \mathcal{S}_{\text{trivariate}}^{\text{(mildly
connected)}}\left(  \tau\right)  \right]  \right)  _{\text{correlated}%
}\overset{\tau\rightarrow\infty}{\sim}r_{\text{trivariate}}^{\text{(mildly
connected)}}\left(  \rho\right)  \left(  \exp\left[  \mathcal{S}%
_{\text{trivariate}}^{\text{(mildly connected)}}\left(  \tau\right)  \right]
\right)  _{\text{uncorrelated}}\text{,} \label{cacchio2}%
\end{equation}
where%
\begin{equation}
r_{\text{trivariate}}^{\text{(mildly connected)}}\left(  \rho\right)
=\sqrt{\frac{3\left(  1-2\rho^{2}\right)  }{3-4\rho}}\text{.}%
\end{equation}
The function $r_{\text{trivariate}}^{\text{(mildly connected)}}\left(
\rho\right)  $ exhibits non-monotonic behavior in the correlation coefficient
$\rho$ and assumes a value equal to zero at the extrema of the allowed range
$\rho\in\left(  -\frac{\sqrt{2}}{2},\frac{\sqrt{2}}{2}\right)  $. On the other
hand, for closed configurations (i.e. bivariate and trivariate models with all
microscopic variables correlated with each other) the complexity ratio between
correlated and uncorrelated cases exhibits monotonic behavior in the
correlation coefficient $\rho$. For instance, in the fully connected bivariate
Gaussian case, it was determined that,%
\begin{equation}
\left(  \exp\left[  \mathcal{S}_{\text{bivariate}}^{\text{(fully connected)}%
}\left(  \tau\right)  \right]  \right)  _{\text{correlated}}\overset
{\tau\rightarrow\infty}{\sim}r_{\text{bivariate}}^{\text{(fully connected)}%
}\left(  \rho\right)  \left(  \exp\left[  \mathcal{S}_{\text{bivariate}%
}^{\text{(fully connected)}}\left(  \tau\right)  \right]  \right)
_{\text{uncorrelated}}\text{,} \label{cacchio3}%
\end{equation}
where%
\begin{equation}
r_{\text{bivariate}}^{\text{(fully connected)}}\left(  \rho\right)
=\sqrt{1+\rho}\text{.}%
\end{equation}
On the other hand, in the fully connected trivariate Gaussian case, it can be
shown that%
\begin{equation}
\left(  \exp\left[  \mathcal{S}_{\text{trivariate}}^{\text{(fully connected)}%
}\left(  \tau\right)  \right]  \right)  _{\text{correlated}}\overset
{\tau\rightarrow\infty}{\sim}r_{\text{trivariate}}^{\text{(fully connected)}%
}\left(  \rho\right)  \left(  \exp\left[  \mathcal{S}_{\text{trivariate}%
}^{\text{(fully connected)}}\left(  \tau\right)  \right]  \right)
_{\text{uncorrelated}}\text{,} \label{cacchio}%
\end{equation}
where%
\begin{equation}
r_{\text{trivariate}}^{\text{(fully connected)}}\left(  \rho\right)
=\sqrt{1+2\rho}\text{.}%
\end{equation}
It is evident that in the fully connected bivariate and trivariate cases, the
ratios $r_{\text{bivariate}}^{\text{(fully connected)}}\left(  \rho\right)  $
and $r_{\text{trivariate}}^{\text{(fully connected)}}\left(  \rho\right)  $
both exhibit monotonic behavior in $\rho$ over the open intervals $\left(
-1,1\right)  $ and $\left(  -\frac{1}{2},1\right)  $, respectively. By
contrast, in the mildly connected trivariate case depicted in Eq.
(\ref{cacchio2}), a peak in the function $r_{\text{trivariate}}^{\text{(mildly
connected)}}\left(  \rho\right)  $ is observed at $\rho_{\text{peak}}=0.5$
$\geq0$.

We recall that in an antiferromagnetic triangular Ising model with coupling
between neighboring spins equal to $J=-1$, any three neighboring spins are
frustrated \cite{mackay03,landau05}. The frustration arises from the inability
of the spin system to find an energetically favorable ordered state. For the
sake of reasoning, assume that one spin is in the $+1$ state. Then, it is
energetically favorable for its immediate neighbors to be in the opposite
state. However, because of the geometry and/or interactions between the spins,
it is impossible to find an energetically optimal configuration in the case of
an antiferromagnetic triangular Ising model. At best, one can only have two
out of three favorable couplings. Furthermore, when the system is frustrated,
the absence of an ordered state can be described in terms of the absence of a
peak in both the standard deviation of the energy and the heat capacity of the
system as a function of its temperature. Instead, a peak in such
thermodynamical quantities is present when considering a ferromagnetic
triangular Ising model with $J=+1$. Within the IGAC, one would desire a
configuration of minimum complexity in order to make reliable macroscopic
predictions. This requirement is the analogue of the ideal scenario of minimum
energy spin configurations in statistical physics
\cite{mackay03,landau05,sadoc06}. Our results in Eqs. (\ref{cacchio2}) and
(\ref{cacchio}) exhibit a dramatically distinct behavior between the mildly
connected and the fully connected trivariate Gaussian configurations. This
behavior is due to the fact that when carrying out statistical inferences with
positively correlated Gaussian random variables, the system seems to
frustrated in the fully connected case. This happens because the maximum
entropy favorable scenario appears to be incompatible with the ideal scenario
of minimum complexity.\textbf{ }Certain lattice configurations in the presence
of correlations are not especially favorable from a statistical inference
perspective of minimum complexity, just like certain spin configurations are
not particularly favorable from an energy standpoint.

Based on these findings, it was argued in Ref. \cite{felice14} that the
impossibility of attaining the most favorable configuration for certain
correlational structures among microscopic degrees of freedom (from an
entropic inference viewpoint) leads to an information geometric analogue of
the frustration effect that occurs in statistical physics in the presence of
loops \cite{sadoc06}.

\section{Conclusions}

In this manuscript, we presented a brief survey of the main results uncovered
by the authors and collaborators within the framework of the IGAC over the
past decade. As pointed out in the Introductory Background, for the sake of
brevity, we have omitted mathematical details in our main presentation.
However, for the sake of self-consistency, we have added a number of
Appendices covering the basic mathematical details necessary to critically
follow the content of the manuscript. For an extended review with more
mathematical details and physical interpretations on the IGAC, we refer to the
recent work appearing in Ref. \cite{ali17}.

We provided here several illustrative examples of entropic dynamical models
employed to infer macroscopic predictions when only limited information of the
microscopic nature of a system is available. In the first example, we
considered the IGAC applied to a high-dimensional Gaussian statistical model.
In particular, we reported the scaling of the scalar curvature with the
microscopic degrees of freedom of the system in Eq. (\ref{curvature18}), the
asymptotic temporal linear growth of the IGE in\ Eq. (\ref{entropy18}), and,
finally, the asymptotic exponential growth of the Jacobi vector field
intensity on such a curved manifold in Eq. (\ref{jacobi18}). In the second
example, we studied the IGAC of a low-dimensional correlated Gaussian
statistical model and showed that, compared to the uncorrelated scenario where
correlations among microscopic degrees of freedom are absent, the IGE
decreases. This decrease in quantified in terms of a monotonic decreasing
compression factor that depends on the correlation coefficient. This finding
appears in Eq. (\ref{fuckyou}). In the third example, we investigated the IGAC
of a set of uncoupled and anisotropic inverted harmonic oscillators. In
particular, we demonstrated the asymptotic temporal growth of the IGE with
proportionality constant given by the sum of all the frequencies of the
oscillators. This finding is reported in\ Eq. (\ref{Fin}). In the fourth
example, we analyzed the IGAC of integrable and chaotic quantum spin chains.
Specifically, we found that the IGE exhibits asymptotic temporal logarithmic
and linear growth, respectively. The former and latter results appear in Eqs.
(\ref{sintegrable}) and (\ref{schaos}), respectively. In the fifth example, we
studied the information geometric complexity reduction in the presence of a
statistical embedding of a lower-dimensional statistical manifold in a
higher-dimensional one. The observed power law decay of the IGE in terms of
the correlation coefficients that specify the correlational structure that
characterizes the statistical embedding is reported in\ Eq. (\ref{peppino}).
In the sixth illustrative example, we investigated the IGAC applied to a
scattering process between two Gaussian wave packets where quantum
entanglement is generated. Conjecturing that the pre- and post-collisional
quantum dynamical scenarios related to an elastic head-on collision are
macroscopic manifestations emerging from microscopic statistical structures,
the IGAC allows to link the behavior of the complexity of motion to the
presence of entanglement. Equation (\ref{family}) illustrates this statement.
In the seventh example, we compared the IGAC applied to a three-dimensional
Gaussian statistical model with that of a two-dimensional Gaussian model
obtained from the former model upon exploitation of a suitable macroscopic
information constraint (see Eq. (\ref{macro})) that resembles the Heisenberg
uncertainty relation. In particular, we found that the IGE\ of the
lower-dimensional statistical model is softened with respect to the
higher-dimensional one. This finding appears in\ Eq. (\ref{soft}). Finally, in
the eighth example, we analyzed the information geometric complexity behavior
of topologically distinct statistical correlational structures for underlying
curved statistical manifolds of different dimensionality. The outcomes of this
particular investigation are presented in Eqs. (\ref{cacchio2}),
(\ref{cacchio3}), and (\ref{cacchio}).

We are aware of several unresolved issues within the IGAC. In what follows, we
outline in a more systematic fashion a number of strengths and weaknesses of
the IGAC theoretical scheme.

\emph{Strengths}.\emph{ }IGAC is characterized by a number of very convenient features:

\begin{description}
\item[i)] No arbitrariness or lack of explanation of how macrostates of a
system leading to the formation of geodesic paths on the curved statistical
manifold is present. Within the IGAC, the transition (that is, the updating)
from an initial macrostate to a final macrostate occurs by navigating through
a continuous sequence of intermediate macrostates chosen by maximizing the
relative entropy between any two consecutive intermediate macrostates subject
to the available information constraints;

\item[ii)] All the dynamical information is collected into a single geometric
quantity where all the available symmetries are retained: the curved
statistical manifold on which the geodesic flow is induced. For instance, the
sensitive dependence of trajectories on initial conditions can be analyzed
from the geodesic deviation equation. Furthermore, the non-integrability
(chaoticity) of the system can be studied by investigating the existence
(absence) of Killing tensors on the curved manifold;

\item[iii)] IGAC\ offers a unifying theoretical setting wherein both curvature
and entropic indicators of complexity are available;

\item[iv)] IGAC represents a convenient platform for enhancing our
comprehension of the role played by statistical curvature in modelling
realistic processes by linking it to conventionally accepted quantities,
including entropy;

\item[v)] From a more foundational perspective, provided that the true degrees
of freedom of the system are identified, IGAC\ presents a serious opportunity
to uncover deep insights into the foundations of modelling and inductive
reasoning together with the relationship to each other.
\end{description}

\emph{Weaknesses}.\emph{ }Despite its strengths, the current version of the
IGAC needs to be improved since it exhibits several weak points, including:

\begin{description}
\item[i)] IGE lacks a detailed comparison with other entropic complexity
indicators of geometric flavor;

\item[ii)] Despite the interpretational power of the Riemannian geometrization
of dynamics, the integration of geodesic equations together with computations
involving curvatures and Jacobi fields can become quite challenging,
especially for higher-dimensional statistical manifolds lacking any particular
degree of symmetry;

\item[iii)] There is no fully-developed quantum mechanical IGAC framework
suitable for characterizing the complexity of quantum evolution;

\item[iv)] IGAC lacks experimental evidence in support of theoretical
macroscopic predictions advanced within its setting;

\item[v)] General results with a wide range of applicability are absent. Most
macroscopic predictions are limited to specific classes of physical systems in
the presence of very peculiar functional forms of relevant available information.
\end{description}

Despite these weaknesses, we are truly gratified\ that the IGAC is gradually
gaining attention within the scientific community. Indeed, there seems to be
an increasing number of scientists who either actively make use of, or whose
work is related to, the IGAC \cite{r20,peng, peng2, r1, r2, r3, FMP, r5, r6,
r7, r8, r9, r10, r11, r12, r13, r14, r15, r16, r17,r19, r18,r18a}.

In conclusion, we emphasize that it was not our intention to report in this
manuscript all the available scientific investigations on complexity based
upon the information geometric approach. Instead, we limited our presentation
to the findings uncovered within the framework of the so-called
IGAC\ theoretical framework. For an overview of various methods of information
geometry used to quantify the complexity of physical systems in both classical
and quantum settings, we refer to the recent review article in Ref. \cite{r20}
and to the works cited therein.

\begin{acknowledgments}
C. C. acknowledges the hospitality of the United States Air Force Research
Laboratory in Rome (New York) where part of his initial contribution to this
work was completed. Finally, the authors are grateful to Dr. Domenico Felice
and Dr. Daniel Stevenson for helpful comments.
\end{acknowledgments}

\bigskip\pagebreak

\begin{center}
\textbf{Conflict of Interest}
\end{center}

The authors declare that there is no conflict of interest regarding the
publication of this paper.

\pagebreak

\appendix

\section{Curvature}

In this Appendix, we review some basic mathematical details on the concept of
curvature. Recall that an $n$-dimensional $%
\mathbb{C}
^{\infty}$ differentiable manifold is a set of points $\mathcal{M}$ that is
endowed with coordinate systems $\mathcal{C}_{\mathcal{M}}$ and fulfills the
following two conditions: 1) each element $c\in\mathcal{C}_{\mathcal{M}}$ is a
one-to-one mapping from $\mathcal{M}$ to some open subset of $%
\mathbb{R}
^{n}$; 2) for all $c\in\mathcal{C}_{\mathcal{M}}$, given any one-to-one
mapping $\eta$ from $\mathcal{M}$ to $%
\mathbb{R}
^{n}$, we find that $\eta\in\mathcal{C}_{\mathcal{M}}\Leftrightarrow\eta\circ
c^{-1}$ is a $%
\mathbb{C}
^{\infty}$ diffeomorphism.

In this manuscript, the points of $\mathcal{M}$ are probability distributions.
Moreover, we take into consideration Riemannian manifolds $\left(
\mathcal{M}\text{, }g\right)  $. The structure of $\mathcal{M}$ as a manifold
does not naturally determine the Riemannian metric $g$. Formally, an infinite
number of Riemannian metrics on $\mathcal{M}$ can be considered. A key working
assumption in the information geometry framework is the choice of the
Fisher-Rao information metric as the metric that underlies the Riemannian
geometry of probability distributions \cite{amari,fisher,rao},%
\begin{equation}
g_{\mu\nu}\left(  \theta\right)  \overset{\text{def}}{=}\int p\left(
x|\theta\right)  \partial_{\mu}\log p\left(  x|\theta\right)  \partial_{\nu
}\log p\left(  x|\theta\right)  dx\text{,}\label{FR}%
\end{equation}
with $\mu$, $\nu=1$,..., $n$ for an $n$-dimensional manifold and
$\partial_{\mu}\overset{\text{def}}{=}\frac{\partial}{\partial\theta^{\mu}}$.
The quantity $x$ in Eq. (\ref{FR}) labels the microstates of the system. The
most compelling support of the choice of the information metric comes from
Cencov's characterization theorem \cite{cencov}. In this theorem, Cencov
proves that the information metric is the only Riemannian metric, up to any
arbitrary constant scale factor, that remains invariant under a family of
probabilistically meaningful mappings \textbf{(}named congruent
embeddings\textbf{)} by Markov morphism \cite{cencov, campbell}.

Once the Fisher-Rao information metric $g_{\mu\nu}\left(  \theta\right)  $ in
Eq. (\ref{FR})\ has been introduced, we can use standard differential geometry
methods applied to the space of probability distributions to characterize the
geometric properties of a curved statistical manifold $\mathcal{M}_{s}$. For
instance, the Ricci scalar curvature $\mathcal{R}_{\mathcal{M}_{s}}$ is given
by \cite{weinberg},%
\begin{equation}
\mathcal{R}_{\mathcal{M}_{s}}\overset{\text{def}}{=}g^{\mu\nu}\mathcal{R}%
_{\mu\nu}\text{,} \label{ricci-scalar}%
\end{equation}
where $g^{\mu\nu}g_{\nu\rho}=\delta_{\rho}^{\mu}$ so that $g^{\mu\nu}=\left(
g_{\mu\nu}\right)  ^{-1}$. The Ricci tensor $\mathcal{R}_{\mu\nu}$ in Eq.
(\ref{ricci-scalar}) is defined as \cite{weinberg},%
\begin{equation}
\mathcal{R}_{\mu\nu}\overset{\text{def}}{=}\partial_{\gamma}\Gamma_{\mu\nu
}^{\gamma}-\partial_{\nu}\Gamma_{\mu\lambda}^{\lambda}+\Gamma_{\mu\nu}%
^{\gamma}\Gamma_{\gamma\eta}^{\eta}-\Gamma_{\mu\gamma}^{\eta}\Gamma_{\nu\eta
}^{\gamma}\text{.} \label{ricci-tensor}%
\end{equation}
The Christoffel connection coefficients $\Gamma_{\mu\nu}^{\rho}$ that appear
in the Ricci tensor in Eq. (\ref{ricci-tensor}) are defined in the standard
manner as \cite{weinberg},
\begin{equation}
\Gamma_{\mu\nu}^{\rho}\overset{\text{def}}{=}\frac{1}{2}g^{\rho\sigma}\left(
\partial_{\mu}g_{\sigma\nu}+\partial_{\nu}g_{\mu\sigma}-\partial_{\sigma
}g_{\mu\nu}\right)  \text{.} \label{connection}%
\end{equation}

We remark\textbf{ }at this point that a geodesic on a $n$-dimensional curved
statistical manifold $\mathcal{M}_{s}$ represents the maximum probability path
a complex dynamical system explores in its evolution between initial and final
macrostates $\theta_{\text{initial}}$ and $\theta_{\text{final}}$,
respectively. Each point of the geodesic represents a macrostate parametrized
by the macroscopic dynamical variables $\theta=\left(  \theta^{1}\text{,...,
}\theta^{n}\right)  $ defining the macrostate of the system. In the framework
of ED, each component $\theta^{j}$ with $j=1$,..., $n$ is a solution of the
geodesic equation \cite{catichaED},%
\begin{equation}
\frac{d^{2}\theta^{k}}{d\tau^{2}}+\Gamma_{lm}^{k}\frac{d\theta^{l}}{d\tau
}\frac{d\theta^{m}}{d\tau}=0\text{.}%
\end{equation}
Furthermore, as stated earlier, each macrostate $\theta$ is in a one-to-one
correspondence with the probability distribution $p\left(  x|\theta\right)  $.
This is a distribution of the microstates $x$. It is also convenient to
observe that the scalar curvature $\mathcal{R}_{\mathcal{M}_{s}}$ can be
expressed as the sum of all sectional curvatures $\mathcal{K}\left(  e_{\rho
}\text{, }e_{\sigma}\right)  $ of planes spanned by pairs of orthonormal basis
elements $\left\{  e_{\rho}=\partial_{\theta_{\rho}(p)}\right\}  $ of the
tangent space $T_{p}\mathcal{M}_{s}$ with $p\in\mathcal{M}_{s}$,%
\begin{equation}
\mathcal{R}_{\mathcal{M}_{s}}\overset{\text{def}}{=}\mathcal{R}_{\text{
}\alpha}^{\alpha}\overset{\text{def}}{=}\sum_{\rho\neq\sigma}\mathcal{K}%
\left(  e_{\rho}\text{, }e_{\sigma}\right)  \text{,} \label{Ricci}%
\end{equation}
where $\mathcal{K}\left(  a\text{, }b\right)  $ is defined as \cite{weinberg},%
\begin{equation}
\mathcal{K}\left(  a\text{, }b\right)  \overset{\text{def}}{=}\frac
{\mathcal{R}_{\mu\nu\rho\sigma}a^{\mu}b^{\nu}a^{\rho}b^{\sigma}}{\left(
g_{\mu\sigma}g_{\nu\rho}-g_{\mu\rho}g_{\nu\sigma}\right)  a^{\mu}b^{\nu
}a^{\rho}b^{\sigma}}\text{,} \label{sectionK}%
\end{equation}
with,%
\begin{equation}
a\overset{\text{def}}{=}\sum_{\rho}\left\langle a\text{, }e^{\rho
}\right\rangle e_{\rho}\text{, }b\overset{\text{def}}{=}\sum_{\rho
}\left\langle b\text{, }e^{\rho}\right\rangle e_{\rho}\text{, and
}\left\langle e_{\rho}\text{, }e^{\sigma}\right\rangle \overset{\text{def}}%
{=}\delta_{\rho}^{\sigma}\text{.}%
\end{equation}
Note that the sectional curvatures $\mathcal{K}\left(  e_{\rho}\text{,
}e_{\sigma}\right)  $ completely determine the Riemann curvature tensor
$\mathcal{R}_{\alpha\beta\rho\sigma}$ where \cite{weinberg},%
\begin{equation}
\mathcal{R}^{\alpha}\,_{\beta\rho\sigma}\overset{\text{def}}{=}g^{\alpha
\gamma}\mathcal{R}_{\gamma\beta\rho\sigma}\overset{\text{def}}{=}%
\partial_{\sigma}\Gamma_{\text{ \ }\beta\rho}^{\alpha}-\partial_{\rho}%
\Gamma_{\text{ \ }\beta\sigma}^{\alpha}+\Gamma^{\alpha}\,_{\lambda\sigma
}\Gamma^{\lambda}\,_{\beta\rho}-\Gamma^{\alpha}\,_{\lambda\rho}\Gamma
^{\lambda}\,_{\beta\sigma}\text{.}%
\end{equation}

We point out that the negativity of the Ricci curvature $\mathcal{R}%
_{\mathcal{M}_{s}}$ is a strong (that is, sufficient but not necessary)
criterion of dynamical instability and that the compactness of the manifold
$\mathcal{M}_{s}$\ \ is necessary in order to handle true chaotic (i.e.,
temporally complex) dynamical systems. Specifically, we observe that from Eq.
(\ref{Ricci}) the negativity of $\mathcal{R}_{\mathcal{M}_{s}}$ implies that
negative principal curvatures (that is, extrema of sectional curvatures)
dominate over positive ones. Therefore, the negativity of the Ricci scalar is
only a sufficient but not necessary condition for local instability of
geodesic flows on curved statistical manifolds. Given these observations, we
reach the conclusion that the negativity of the Ricci scalar curvature
provides a strong\textit{\ }criterion of local instability. We also point out
the possible occurrence of scenarios where negative sectional curvatures are
present, but the positive ones prevail in the sum in Eq. (\ref{Ricci}) so that
$\mathcal{R}_{\mathcal{M}_{s}}$ is non-negative despite the instability in the
flow in those directions. In summary, to properly characterize the chaoticity
(i.e., temporal complexity) of complex dynamical systems, the signs of the
sectional curvatures are of primary importance. For further mathematical
details on the notion of curvature in differential geometry, we refer to Ref.
\cite{lee}.

\section{Information Geometric Entropy}

In this Appendix, we present the concept of the IGE within the IGAC
theoretical setting. Assume that the points $\left\{  p\left(  x;\theta
\right)  \right\}  $ of an $n$-dimensional curved statistical manifold
$\mathcal{M}_{s}$ are parametrized by means of $n$ \emph{real} valued
variables $\left(  \theta^{1}\text{,..., }\theta^{n}\right)  $,
\begin{equation}
\mathcal{M}_{s}\overset{\text{def}}{=}\left\{  p\left(  x;\theta\right)
:\theta=\left(  \theta^{1}\text{,..., }\theta^{n}\right)  \in\mathcal{D}%
_{\boldsymbol{\theta}}^{\left(  \text{tot}\right)  }\right\}  \text{.}
\label{smanifold}%
\end{equation}
We remark that the microvariables $x$ in Eq. (\ref{smanifold}) belong to the
microspace $\mathcal{X}$ while the macrovariables $\theta$ in Eq.
(\ref{smanifold}) are elements of the parameter space $\mathcal{D}%
_{\boldsymbol{\theta}}^{\left(  \text{tot}\right)  }$ given by,
\begin{equation}
\mathcal{D}_{\boldsymbol{\theta}}^{\left(  \text{tot}\right)  }\overset
{\text{def}}{=}{\bigotimes\limits_{j=1}^{n}}\mathcal{I}_{\theta^{j}}=\left(
\mathcal{I}_{\theta^{1}}\otimes\mathcal{I}_{\theta^{2}}\text{...}%
\otimes\mathcal{I}_{\theta^{n}}\right)  \subseteq\mathbb{R}^{n}\text{.}
\label{dtot}%
\end{equation}
The quantity $\mathcal{I}_{\theta^{j}}$ with $1\leq j\leq n$ in Eq.
(\ref{dtot}) is a subset of $\mathbb{R}^{n}$ and denotes the entire range of
allowable values for the statistical macrovariables $\theta^{j}$. The
information geometric entropy is a proposed measure of temporal complexity of
geodesic paths within the IGAC. The IGE\ is defined as,
\begin{equation}
\mathcal{S}_{\mathcal{M}_{s}}\left(  \tau\right)  \overset{\text{def}}{=}%
\log\widetilde{vol}\left[  \mathcal{D}_{\boldsymbol{\theta}}\left(
\tau\right)  \right]  \text{,} \label{IGE}%
\end{equation}
where the average dynamical statistical volume\textbf{\ }$\widetilde
{vol}\left[  \mathcal{D}_{\boldsymbol{\theta}}\left(  \tau\right)  \right]  $
is given by,
\begin{equation}
\widetilde{vol}\left[  \mathcal{D}_{\boldsymbol{\theta}}\left(  \tau\right)
\right]  \overset{\text{def}}{=}\frac{1}{\tau}\int_{0}^{\tau}vol\left[
\mathcal{D}_{\boldsymbol{\theta}}\left(  \tau^{\prime}\right)  \right]
d\tau^{\prime}\text{.} \label{rhs}%
\end{equation}
Observe that the operation of temporal average is denoted with the tilde
symbol in Eq. (\ref{rhs}). Moreover, the volume\textbf{\ }$vol\left[
\mathcal{D}_{\boldsymbol{\theta}}\left(  \tau^{\prime}\right)  \right]
$\textbf{\ }in the RHS of Eq. (\ref{rhs}) is defined as,
\begin{equation}
vol\left[  \mathcal{D}_{\boldsymbol{\theta}}\left(  \tau^{\prime}\right)
\right]  \overset{\text{def}}{=}\int_{\mathcal{D}_{\boldsymbol{\theta}}\left(
\tau^{\prime}\right)  }\rho\left(  \theta^{1}\text{,..., }\theta^{n}\right)
d^{n}\theta\text{,} \label{v}%
\end{equation}
where $\rho\left(  \theta^{1}\text{,..., }\theta^{n}\right)  $ is the
so-called Fisher density and equals the square root of the determinant
$g\left(  \theta\right)  $ of the Fisher-Rao information metric tensor
$g_{\mu\nu}\left(  \theta\right)  $,
\begin{equation}
\rho\left(  \theta^{1}\text{,..., }\theta^{n}\right)  \overset{\text{def}}%
{=}\sqrt{g\left(  \theta\right)  }\text{.}%
\end{equation}
We point out that the expression of $vol\left[  \mathcal{D}%
_{\boldsymbol{\theta}}\left(  \tau^{\prime}\right)  \right]  $ in Eq.
(\ref{v}) can become more transparent for statistical manifolds with
information metric tensor whose determinant can be factorized in the following
manner,
\begin{equation}
g\left(  \theta\right)  =g\left(  \theta^{1}\text{,..., }\theta^{n}\right)
={\prod\limits_{j=1}^{n}}g_{j}\left(  \theta^{j}\right)  \text{.}%
\end{equation}
With the aid of the factorized determinant, the IGE in Eq. (\ref{IGE}) can be
recast as
\begin{equation}
\mathcal{S}_{\mathcal{M}_{s}}\left(  \tau\right)  =\log\left\{  \frac{1}{\tau
}\int_{0}^{\tau}\left[  {\prod\limits_{j=1}^{n}}\left(  \int_{\tau_{0}}%
^{\tau_{0}+\tau^{\prime}}\sqrt{g_{j}\left[  \theta^{j}\left(  \xi\right)
\right]  }\frac{d\theta^{j}}{d\xi}d\xi\right)  \right]  d\tau^{\prime
}\right\}  \text{.} \label{IGEmod}%
\end{equation}
We also emphasize that within the IGAC, the leading asymptotic behavior of
$\mathcal{S}_{\mathcal{M}_{s}}\left(  \tau\right)  $ is employed to quantify
the complexity of the statistical models being investigated. For this reason,
it is customary to consider the quantity
\begin{equation}
\mathcal{S}_{\mathcal{M}_{s}}^{\left(  \text{asymptotic}\right)  }\left(
\tau\right)  \sim\lim_{\tau\rightarrow\infty}\left[  \mathcal{S}%
_{\mathcal{M}_{s}}\left(  \tau\right)  \right]  \text{,}%
\end{equation}
that is to say, the leading asymptotic term in the IGE expression. The
integration space $\mathcal{D}_{\theta}\left(  \tau^{\prime}\right)  $ in Eq.
(\ref{v}) is defined by,
\begin{equation}
\mathcal{D}_{\boldsymbol{\theta}}\left(  \tau^{\prime}\right)  \overset
{\text{def}}{=}\left\{  \theta:\theta^{j}\left(  \tau_{0}\right)  \leq
\theta^{j}\leq\theta^{j}\left(  \tau_{0}+\tau^{\prime}\right)  \right\}
\text{,} \label{is}%
\end{equation}
where $\theta^{j}=\theta^{j}\left(  \xi\right)  $ with $\tau_{0}\leq\xi
\leq\tau_{0}+\tau^{\prime}$ and $\tau_{0}$ denoting the initial value of the
affine parameter $\xi$ such that,
\begin{equation}
\frac{d^{2}\theta^{j}\left(  \xi\right)  }{d\xi^{2}}+\Gamma_{ik}^{j}%
\frac{d\theta^{i}}{d\xi}\frac{d\theta^{k}}{d\xi}=0\text{.} \label{ge}%
\end{equation}
The integration domain $\mathcal{D}_{\boldsymbol{\theta}}\left(  \tau^{\prime
}\right)  $ in Eq. (\ref{is}) is an $n$-dimensional subspace of $\mathcal{D}%
_{\boldsymbol{\theta}}^{\left(  \text{tot}\right)  }$ whose elements are
$n$-dimensional macrovariables $\left\{  \theta\right\}  $ with components
$\theta^{j}$ bounded by given limits of integration $\theta^{j}\left(
\tau_{0}\right)  $ and $\theta^{j}\left(  \tau_{0}+\tau^{\prime}\right)  $.
The integration of the $n$-coupled nonlinear second order ordinary
differential equations in Eq. (\ref{ge}) specifies the temporal functional
form of such limits.

The IGE at a certain instant is essentially the logarithm of the volume of the
effective parameter space explored by the system at that very instant. In
order to average out the possibly highly complex fine details of the entropic
dynamical description of the system on the curved statistical manifold, the
temporal average has been taken into consideration. Furthermore, to eliminate
the consequences of transient effects which enter the computation of the
expected value of the volume of the effective parameter space, the long-term
asymptotic temporal behavior is considered to conveniently describe the
selected dynamical complexity indicator. In summary, the IGE is constructed to
furnish an asymptotic coarse-grained inferential description of the complex
dynamics of a system in the presence of only partial knowledge. For further
details on the IGE, we refer to Refs. \cite{cafaroamc10,ali17}.

\section{Jacobi Fields}

In this Appendix, we review some basic mathematical details on the concept of
Jacobi vector fields. The investigation of the instability of natural motions
by way of the instability of geodesics on a suitable curved manifold is
especially advantageous within the Riemannian geometrization of dynamics. In
particular, the so-called Jacobi-Levi-Civita (JLC) equation for geodesic
spread is a very powerful mathematical tool employed to study the
stability/instability of a geodesic flow. This equation is a familiar quantity
both in theoretical physics (in General Relativity, for instance) and in
Riemannian geometry. The JLC equation covariantly describes how neighboring
geodesics locally scatter. More specifically, the JLC equation connects
curvature properties of the ambient manifold to the stability/instability of a
geodesic flow. It paves the way to a wide and largely unexplored field of
investigation that concerns the links among geometry, topology and geodesic
instability, and therefore to chaoticity and complexity. To the best of our
knowledge, the use the JLC equation in the framework of information geometry
appeared originally in Ref. \cite{cafaropd07}.

Let us consider two neighboring geodesic paths $\theta^{\alpha}\left(
\tau\right)  $ and $\theta^{\alpha}\left(  \tau\right)  +\delta\theta^{\alpha
}\left(  \tau\right)  $, with $\tau$ denoting the affine parameter, that
satisfy the following geodesic equations of motion,%
\begin{equation}
\frac{d^{2}\theta^{\alpha}}{d\tau^{2}}+\Gamma_{\beta\gamma}^{\alpha}\left(
\theta\right)  \frac{d\theta^{\beta}}{d\tau}\frac{d\theta^{\gamma}}{d\tau
}=0\text{,} \label{ge1}%
\end{equation}
and%
\begin{equation}
\frac{d^{2}\left[  \theta^{\alpha}+\delta\theta^{\alpha}\right]  }{d\tau^{2}%
}+\Gamma_{\beta\gamma}^{\alpha}\left(  \theta+\delta\theta\right)
\frac{d\left[  \theta^{\beta}+\delta\theta^{\beta}\right]  }{d\tau}%
\frac{d\left[  \theta^{\gamma}+\delta\theta^{\gamma}\right]  }{d\tau
}=0\text{,} \label{ge2}%
\end{equation}
respectively. Observing that to first order in $\delta\theta^{\alpha}$,
\begin{equation}
\Gamma_{\beta\gamma}^{\alpha}\left(  \theta+\delta\theta\right)  \approx
\Gamma_{\beta\gamma}^{\alpha}\left(  \theta\right)  +\partial_{\xi}%
\Gamma_{\beta\gamma}^{\alpha}\delta\theta^{\xi}\text{,}%
\end{equation}
after some algebra, to first order in $\delta\theta^{\alpha}$ Eq. (\ref{ge2})
becomes%
\begin{equation}
\frac{d^{2}\theta^{\alpha}}{d\tau^{2}}+\frac{d^{2}\left(  \delta\theta
^{\alpha}\right)  }{d\tau^{2}}+\Gamma_{\beta\gamma}^{\alpha}\left(
\theta\right)  \frac{d\theta^{\beta}}{d\tau}\frac{d\theta^{\gamma}}{d\tau
}+2\Gamma_{\beta\gamma}^{\alpha}\left(  \theta\right)  \frac{d\theta^{\beta}%
}{d\tau}\frac{d\left(  \delta\theta^{\gamma}\right)  }{d\tau}+\partial_{\xi
}\Gamma_{\beta\gamma}^{\alpha}\left(  \theta\right)  \delta\theta^{\xi}%
\frac{d\theta^{\beta}}{d\tau}\frac{d\theta^{\gamma}}{d\tau}=0\text{.}
\label{ge3}%
\end{equation}
The geodesic deviation equation can be obtained by subtracting Eq. (\ref{ge1})
from Eq. (\ref{ge3}),%
\begin{equation}
\frac{d^{2}\left(  \delta\theta^{\alpha}\right)  }{d\tau^{2}}+2\Gamma
_{\beta\gamma}^{\alpha}\left(  \theta\right)  \frac{d\theta^{\beta}}{d\tau
}\frac{d\left(  \delta\theta^{\gamma}\right)  }{d\tau}+\partial_{\xi}%
\Gamma_{\beta\gamma}^{\alpha}\left(  \theta\right)  \delta\theta^{\xi}%
\frac{d\theta^{\beta}}{d\tau}\frac{d\theta^{\gamma}}{d\tau}=0\text{.}
\label{ge4}%
\end{equation}
Equation (\ref{ge4}) can be rewritten in a more convenient form in terms of
covariant derivatives (see Ref. \cite{ohanian}, for instance) along the curve
$\theta^{\alpha}\left(  \tau\right)  $,%
\begin{align}
\frac{D^{2}\left(  \delta\theta^{\alpha}\right)  }{D\tau^{2}}  &  =\frac
{d^{2}\left(  \delta\theta^{\alpha}\right)  }{d\tau^{2}}+\partial_{\beta
}\Gamma_{\rho\sigma}^{\alpha}\frac{d\theta^{\beta}}{d\tau}\delta\theta^{\rho
}\frac{d\theta^{\sigma}}{d\tau}+2\Gamma_{\rho\sigma}^{\alpha}\frac{d\left(
\delta\theta^{\rho}\right)  }{d\tau}\frac{d\theta^{\sigma}}{d\tau}+\nonumber\\
& \nonumber\\
&  -\Gamma_{\rho\sigma}^{\alpha}\Gamma_{\kappa\lambda}^{\sigma}\delta
\theta^{\rho}\frac{d\theta^{\kappa}}{d\tau}\frac{d\theta^{\lambda}}{d\tau
}+\Gamma_{\rho\sigma}^{\alpha}\Gamma_{\kappa\lambda}^{\rho}\delta
\theta^{\kappa}\frac{d\theta^{\lambda}}{d\tau}\frac{d\theta^{\sigma}}{d\tau
}\text{.} \label{ge5}%
\end{align}
Combining Eqs. (\ref{ge4}) and (\ref{ge5}), after some tensor algebra
manipulations, we obtain%
\begin{equation}
\frac{D^{2}\left(  \delta\theta^{\alpha}\right)  }{D\tau^{2}}=\left(
\partial_{\rho}\Gamma_{\eta\sigma}^{\alpha}-\partial_{\eta}\Gamma_{\rho\sigma
}^{\alpha}+\Gamma_{\lambda\sigma}^{\alpha}\Gamma_{\eta\rho}^{\lambda}%
-\Gamma_{\eta\lambda}^{\alpha}\Gamma_{\rho\sigma}^{\lambda}\right)
\delta\theta^{\eta}\frac{d\theta^{\rho}}{d\tau}\frac{d\theta^{\sigma}}{d\tau
}\text{.} \label{ge6}%
\end{equation}
Finally, observing that the Riemannian curvature tensor components
$\mathcal{R}_{\rho\eta\sigma}^{\alpha}$ are given by,
\begin{equation}
\mathcal{R}_{\rho\eta\sigma}^{\alpha}\overset{\text{def}}{=}\partial_{\eta
}\Gamma_{\rho\sigma}^{\alpha}-\partial_{\rho}\Gamma_{\eta\sigma}^{\alpha
}+\Gamma_{\eta\lambda}^{\alpha}\Gamma_{\rho\sigma}^{\lambda}-\Gamma
_{\lambda\sigma}^{\alpha}\Gamma_{\eta\rho}^{\lambda}\text{,}%
\end{equation}
the component form of the geodesic deviation equation becomes%
\begin{equation}
\frac{D^{2}J^{\alpha}}{D\tau^{2}}+\mathcal{R}_{\rho\eta\sigma}^{\alpha}%
\frac{d\theta^{\rho}}{d\tau}J^{\eta}\frac{d\theta^{\sigma}}{d\tau}=0\text{,}
\label{JLC}%
\end{equation}
where $J^{\alpha}\overset{\text{def}}{=}\delta\theta^{\alpha}$ denotes the
$\alpha$-component of the so-called Jacobi vector field \cite{weinberg}.
Equation (\ref{JLC}) is the so-called JLC equation. From the JLC equation in
Eq. (\ref{JLC}), we note that neighboring geodesics accelerate relative to
each other with a rate directly measured by the Riemannian curvature tensor
$R_{\alpha\beta\gamma\delta}$. The quantity $J^{\alpha}$ is given by,%
\begin{equation}
J^{\alpha}=\delta\theta^{\alpha}\overset{\text{def}}{=}\delta_{\phi}%
\theta^{\alpha}=\left(  \frac{\partial\theta^{\alpha}\left(  \tau\text{; }%
\phi\right)  }{\partial\phi}\right)  _{\tau=\text{constant}}\delta\phi\text{,}
\label{jacobi}%
\end{equation}
with $\left\{  \theta^{\mu}\left(  \tau\text{; }\phi\right)  \right\}  $
denoting the one-parameter $\phi$ family of geodesics whose evolution is
described in terms of the affine parameter $\tau$. The Jacobi vector field
intensity $J\left(  \tau\right)  $ is defined as,%
\begin{equation}
J\left(  \tau\right)  \overset{\text{def}}{=}\left(  J^{\alpha}g_{\alpha\beta
}J^{\beta}\right)  ^{\frac{1}{2}}\text{.}%
\end{equation}
Observe that Eq. (\ref{JLC}) yields a system of coupled ordinary differential
equations \emph{linear} in the components of the deviation vector field
but\textit{\ }\emph{nonlinear} in derivatives of the metric. We remark that
although the JLC equation already seems intractable at rather small
dimensions, in the case of isotropic manifolds it reduces to the following
simplified form,%
\begin{equation}
\frac{D^{2}J^{\mu}}{D\tau^{2}}+\mathcal{K}J^{\mu}=0\text{,}
\label{geo-deviation}%
\end{equation}
where $\mathcal{K}$ is the constant value assumed throughout the manifold by
the sectional curvature. In particular, when $\mathcal{K}<0$, unstable
solutions of Eq. (\ref{geo-deviation}) are of the form%
\begin{equation}
J^{\mu}\left(  \tau\right)  =\frac{\omega_{0}^{\mu}}{\sqrt{-\mathcal{K}}}%
\sinh\left(  \sqrt{-\mathcal{K}}\tau\right)  \text{,}%
\end{equation}
assuming that the initial conditions are given by $J^{\mu}\left(  0\right)
=0$ and $\frac{dJ^{\mu}\left(  0\right)  }{d\tau}=\omega^{\mu}\left(
0\right)  =\omega_{0}^{\mu}\neq0$, respectively, for any $1\leq\mu\leq n$ with
$n$ denoting the dimensionality of the underlying curved manifold. For further
details on the JLC equation, we refer to Refs. \cite{weinberg,carmo,ohanian}.

\end{document}